\documentclass[aps,prl,twocolumn,amsmath,amssymb,floatfix,nofootinbib]{revtex4-1}
\usepackage{natbib}
\usepackage{amssymb,amsmath}
\usepackage{graphicx}
\usepackage{verbatim}
\usepackage{color,hyperref}
\definecolor{linkcolor}{rgb}{0,0,0.25}
\hypersetup{
  colorlinks=true,        
  linkcolor=linkcolor,    
  citecolor=linkcolor,    
  filecolor=linkcolor,    
  urlcolor=linkcolor      
}
\newcommand{\ie}{i.e.}

\newcommand{\dd}{\mathrm{d}}

\newcommand{\Eqnname}{Equation}
\newcommand{\equationname}{\Eqnname}

\renewcommand{\figurename}{Figure}

\newcommand{\Myr}{\ensuremath{\,\mathrm{Myr}}}
\newcommand{\Gyr}{\ensuremath{\,\mathrm{Gyr}}}
\newcommand{\kpc}{\ensuremath{\,\mathrm{kpc}}}
\newcommand{\pc}{\ensuremath{\,\mathrm{pc}}}
\newcommand{\kms}{\ensuremath{\,\mathrm{km\ s}^{-1}}}
\newcommand{\msun}{\ensuremath{\,\mathrm{M}_{\odot}}}

\newcommand{\mnras}{Mon. Not. Roy. Astron. Soc.}
\newcommand{\apjs}{Astrophys. J. Supp.}
\newcommand{\aj}{Astron. J.}
\newcommand{\aap}{Astron. \& Astrophys.}
\newcommand{\jcap}{J. Cosmol. Astropart. Phys}
\newcommand{\nar}{New Astron. Rev.}

\begin{document}

\title{Detecting the disruption of dark-matter halos  with stellar streams}

\author{Jo~Bovy}
\affiliation{Department of Astronomy and Astrophysics, University of Toronto, 
  50 St.  George Street, Toronto, ON, M5S 3H4, Canada;
  bovy@astro.utoronto.ca~}

\date{February 10, 2016}

\begin{abstract}
  Narrow stellar streams in the Milky Way halo are uniquely sensitive
  to dark-matter subhalos, but many of these subhalos may be tidally
  disrupted. I calculate the interaction between stellar and
  dark-matter streams using analytical and $N$-body calculations,
  showing that disrupting objects can be detected as low-concentration
  subhalos. Through this effect, we can constrain the lumpiness of the
  halo as well as the orbit and present position of individual
  dark-matter streams. This will have profound implications for the
  formation of halos and for direct and indirect-detection dark-matter
  searches.
\end{abstract}

\maketitle

\emph{Introduction}---One of the key predictions of the cold-dark
matter paradigm is that the extended dark-matter halos of galaxies
contain a large amount of small-scale structure in the form of
subhalos \citep{Springel08a}. At the high-mass end of the subhalo
spectrum, this structure is visible in the form of dwarf galaxies
\citep{Belokurov13a}. But if dark matter is truly cold, the mass
spectrum should extend well below the halo mass scale where baryons
can condense and form stars and below the scales constrained by the
Ly-$\alpha$ forest ($M \gtrsim 3\times10^8\msun$; \cite{Viel13a}). An
observational determination of the subhalo mass spectrum to well below
$10^9\msun$ would provide one of the most important astrophysical
constraints on the nature of dark matter.

In our own Milky Way galaxy, one of the most promising methods for
detecting low-mass dark-matter subhalos is through their effects on
cold stellar streams in the halo
\citep{Johnston02a,Ibata02a,Carlberg12a}. Cold stellar streams form
when a globular cluster in the halo gets tidally disrupted; mass loss
at pericentric passages is deposited into orbits with slightly higher
and lower orbital energies, leading to narrow leading and trailing
arms \citep{Johnston98a}. Many examples of such streams are now known
from wide-area photometric surveys
\citep{Odenkirchen01a,Grillmair06a}. The kinematical coldness of tidal
streams makes them sensitive to the influence of subhalos with masses
$\lesssim10^8\msun$. Dynamical modeling of the smooth stream itself
\citep{Bovy14a,Sanders14a} and of the impact of subhalos
\citep{Erkal15a,Sanders15a} has been shown to be able to detect and
characterize subhalos with masses down to $10^7\msun$ with \emph{Gaia}
and LSST \citep{Erkal15b}.

While many of the dark-matter subhalos are expected to survive as
separate entities, some of them, especially the more massive ones
within a few tens of kpc from the Galactic center, may be in the
process of being tidally disrupted in a similar manner as the globular
clusters \citep{Zemp09a}. If this is the case, a null detection of the
expected cold-dark-matter mass spectrum at $M\lesssim10^9\msun$ could
be misconstrued as evidence against cold dark matter. On the more
positive side, if a significant fraction of the dark matter in the
solar neighborhood is coherent in velocity space, annual modulation in
dark-matter direct-detection experiments may be enhanced
\citep{Kuhlen10a,Lisanti12a} and dark matter in models with high
minimum scattering thresholds would be easier to detect
\citep{Smith01a,Bottino04a}. A determination of the fraction of
dark-matter subhalos that are in the process of tidal disruption is
therefore crucial for future astrophysical and direct-detection
experiments into the nature of dark matter.

In this Letter, I compute the impact of dark-matter streams---formed
from subhalos that are in the process of being tidally disrupted---on
stellar streams. The kinematical coldness of stellar streams makes
them excellent probes of this scenario. The interaction between a
dark-matter and a stellar stream is more extended than that between a
surviving subhalo and a stellar stream, and therefore a larger part of
the stream is affected by the interaction. This implies that
dark-matter streams will most easily be detected in convential
analyses (\ie, which assume a surviving subhalo) as detections with
subhalo parameters that imply anomalously low concentrations. The
extended interaction, however, also causes the impulse
approximation---which is typically accurate for subhalo--stream
interactions---to break down and I demonstrate that large velocity
kicks can occur, even for very extended dark-matter tidal tails, for
which the cross section is high. The breakdown of the impulse
approximation also opens up the possibility that we can infer the
orbit and current location of individual, entirely-dark subhalos.

\begin{figure}
\includegraphics[width=0.45\textwidth]{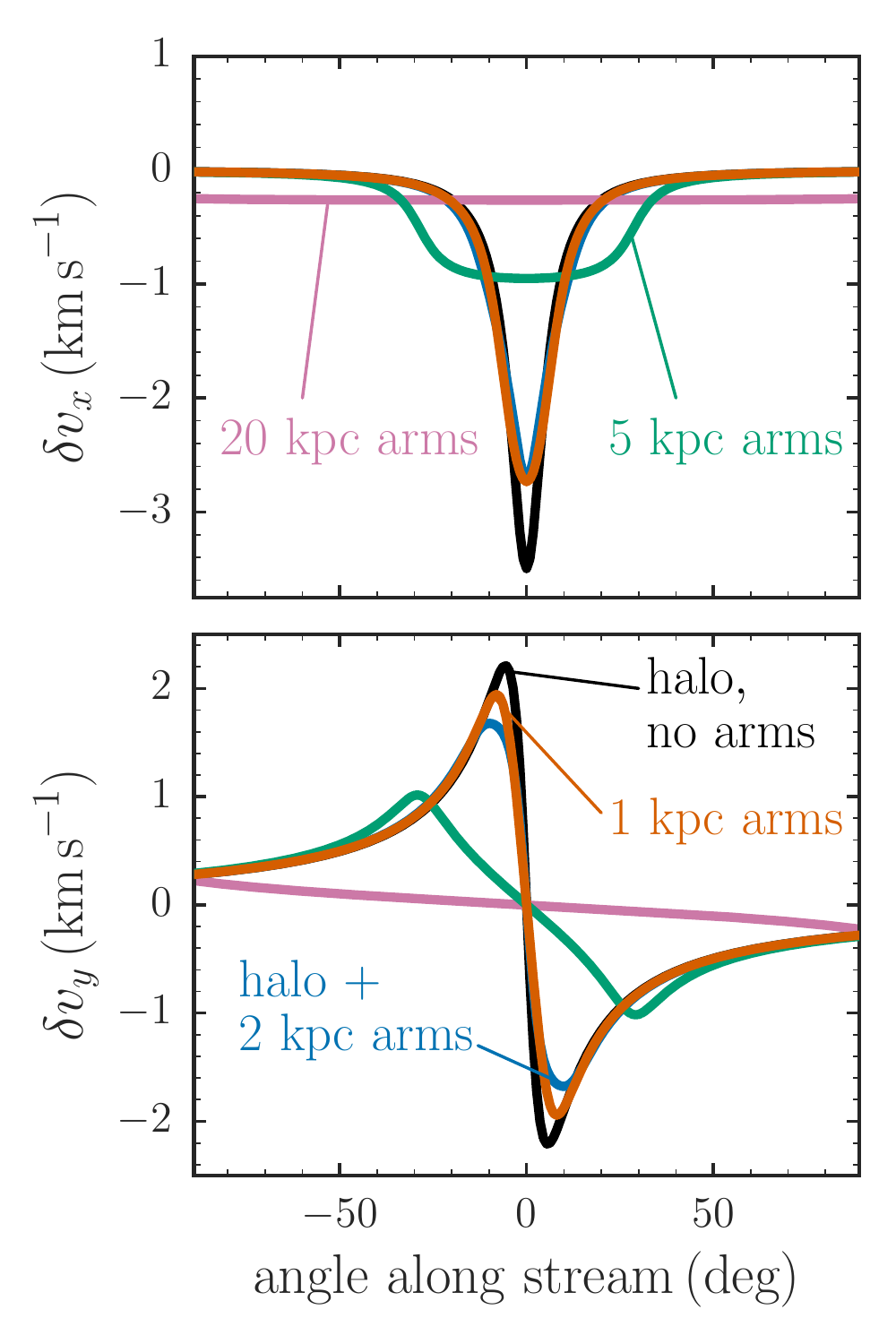}
\caption{Interaction between a stream on a circular orbit at 10\,kpc
  in the $x-y$ plane moving at $220\kms$ anticlockwise with a
  dark-matter stream with a total mass of $10^8\msun$ moving at
  $(0,132,176)\kms$, making a closest approach at $625\pc$ from $(x,y)
  = (10,0)\kpc$ (the setup of Figure 1 in
  Ref. \citep{Sanders15a}). The interaction is computed in the impulse
  approximation of \equationname s~\eqref{eq:straight_impulse},
  assuming uniform $\dd GM / \dd t$ for different stream lengths and
  $r_s(t) = 625\pc$. The curve labeled as `halo, no arms' has all of
  the mass in a single Plummer sphere; that labeled as `halo+$2\kpc$
  arms' has half of the mass in a stream and half in a single Plummer
  halo. An interaction with a stream rather than a surviving subhalo
  has a lower amplitude, but affects a much larger part of the stellar
  stream.\label{fig:straight}}
\end{figure}

\emph{Impulse approximation}---The interaction between a stellar
stream and a surviving dark-matter subhalo is typically well described
using the \emph{impulse approximation} \citep{Sanders15a,Yoon11a}. In
this approximation, both the subhalo and the stream are approximated
as moving on a straight line at the time of closest approach; the
interaction is modeled as an instantaneous velocity kick along the
stellar stream. We can calculate the interaction between a dark-matter
and stellar stream in a similar manner, approximating the dark-matter
stream as a set of Plummer spheres \citep{Plummer11a} for which the
interaction with the stellar stream can be computed
analytically. Using the same setup as in Ref. \citep{Erkal15a}, where
the stellar stream moves along the $y$ axis with velocity $v_y$ and
the dark-matter moves with velocity
$(-w_\perp\,\sin\alpha,w_y,w_\perp\,\cos\alpha)$ through the point of
closest approach at $(b\cos\alpha,0,b\sin\alpha)$, the velocity kicks
along the stream are
\begin{align}\label{eq:straight_impulse}
  \Delta v_y & = -\int \dd t \frac{\dd GM}{\dd t}\frac{w_\perp^2\tilde{y}(t)}{w\left([b^2+r_s^2(t)]w^2+w_\perp^2\tilde{y}(t)^2\right)}\,,\\
  \Delta v_{x/z} & = 2\int \dd t \frac{\dd GM}{\dd t}\frac{bw^2\cos[\sin]\alpha\pm \tilde{y}(t) w_\perp w_\parallel\sin[\cos]\alpha}{w\left([b^2+r_s^2(t)]w^2+w_\perp^2\tilde{y}(t)^2\right)}\,,\nonumber\\
\mathrm{where} &  \ \qquad \ \tilde{y}(t) = y-v_y t\nonumber\,.
\end{align}
In these expressions, $w_\parallel = v_y-w_y$, $w =
\sqrt{w_\perp^2+w_\parallel^2}$, and $\dd GM / \dd t$ is the mass of
the Plummer sphere with scale radius $r_s(t)$ that passes through the
point of closest approach at time $t$; choose the $\sin$ and $\cos$ in
brackets and the minus sign for $\Delta v_z$. This expression assumes
that the velocity kicks arising from different parts of the
dark-matter stream add linearly, which is a good assumption for the
small kicks from $M \lesssim10^9\msun$ subhalos.

As discussed in Ref. \citep{Erkal15a}, the overall amplitude of the
kick of a single Plummer sphere is primarily set by the mass. The
extent $\Delta y$ over which the kicks are significant is set by the
scale radius $r_s$ and the impact parameter $b$. It is clear from the
expressions in \equationname s~\eqref{eq:straight_impulse} that if the
dark-matter stream has a length $L$ that is short compared to $\Delta
y$ in the sense that $v_y L / \sqrt{w_\perp^2+w_y^2} < \Delta y$, then
the interaction is similar to that with a single subhalo. Because the
two stream velocities are generically similar, this is only the case
for dark-matter streams that are not much longer than $r_s$, \ie, very
early in the disruption process. For longer dark-matter streams, the
interaction will be softened as the various parts of the leading and
trailing tails of the dark-matter stream typically produce kicks in
opposite directions. The net effect is to reduce the amplitude of the
velocity kick below that from a surviving subhalo, while
simultaneously acting over a larger part of the stellar stream. This
is illustrated in \figurename~\ref{fig:straight}. It is
straightforward to generalize the impulse approximation in the
previous paragraph to take into account the curved nature of the
stellar stream (cf. Ref. \citep{Sanders15a}). This can be efficiently
done by moving the stream along an orbit to compute the velocities of
the stream segments over the time interval of the interaction.

\begin{figure*}
\includegraphics[width=\textwidth]{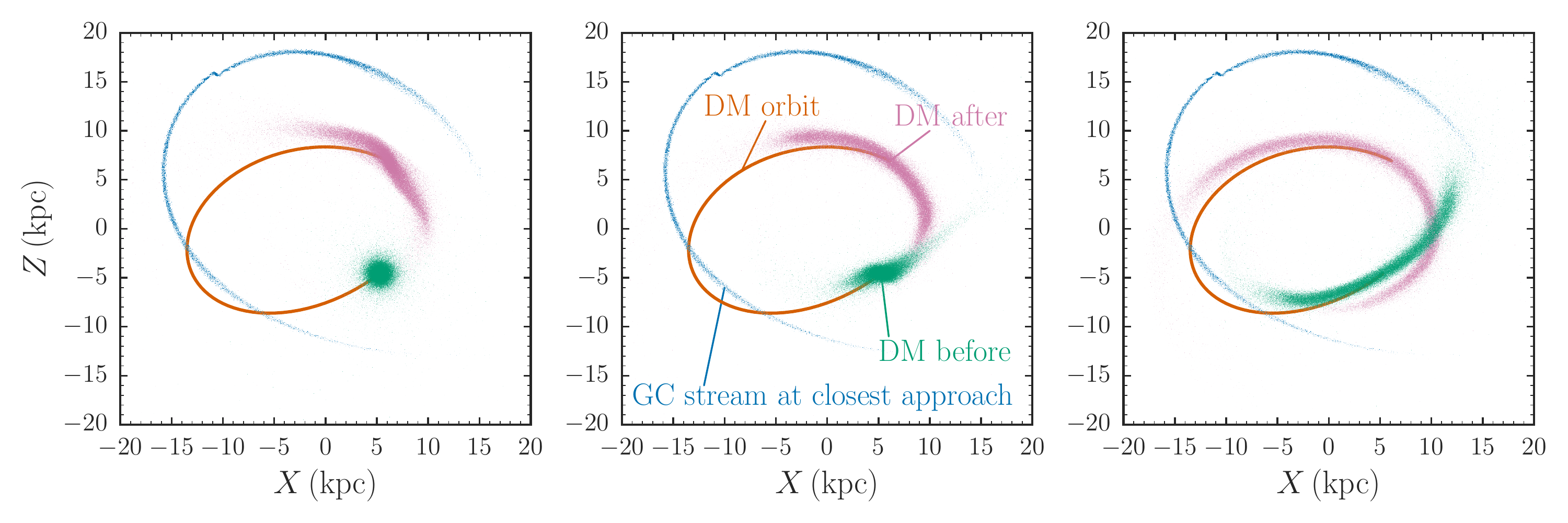}
\caption{$N$-body simulations of the interaction between a dark-matter
  (DM) and globular-cluster (GC) stellar stream. The stellar stream is
  shown at the point of closest approach between the stream and the
  dark-matter progenitor. The dark-matter is displayed $125\pc/(\kms)$
  ($\approx125\Myr$) before and $125\pc/(\kms)$ after the interaction,
  which is the time interval over which the DM and GC streams are
  evolved together. The orbit of the dark-matter progenitor during
  this time is given in red. Three different dark-matter streams are
  generated by letting the dark-matter disrupt for different amounts
  of time. In the simulation on the left, the DM stream is only
  starting to form, in the middle panel a long DM stream is in the
  process of forming, and on the right the DM subhalo is fully
  disrupted, but still forms a coherent stream.\label{fig:nbody}}
\end{figure*}

\emph{$N$-body simulations}---To investigate the interaction between
two streams in further detail, I use $N$-body simulations to compute
the full, non-linear interaction. The setup of these simulations is as
in Ref. \citep{Sanders15a}. A mock stellar stream is generated by
evolving a King cluster \citep{King66a} of $10^5\msun$ with $W_0 = 5$
and a core radius of $13\pc$ represented with $10^5$ particles for
$10.125\kpc/(\kms)$ ($\approx10\Gyr$) in a logarithmic host potential
with a circular velocity of $220\kms$ and a potential flattening of
0.9. This mock stellar stream sustains a direct hit by dark-matter
streams at $(X,Y,Z) = (-13.5,2.84,-1.84)\kpc$ moving at $(v_x,v_y,v_z)
= (6.82,132.77,149.42)\kms$, generated by evolving a Plummer sphere
with $M = 10^8\msun$ and $r_s = 625\pc$ (using $10^5$ particles) for
$125\pc/(\kms)$, $250\pc/(\kms)$, and $500\pc/(\kms)$. The dark-matter
and stellar streams are evolved together for $250\pc/(\kms)$
($\approx250\Myr$) starting $125\pc/(\kms)$ before the direct-impact
time (defined as the time at which the progenitor dark-matter subhalo
would have directly hit the stellar stream), to be able to study the
interaction in a clean manner. All $N$-body simulations are run using
\texttt{gyrfalcON} and \texttt{NEMO}
\citep{Dehnen02a,Teuben95a}. These $N$-body simulations are
demonstrated in \figurename~\ref{fig:nbody}.

\begin{figure*}
\includegraphics[width=\textwidth]{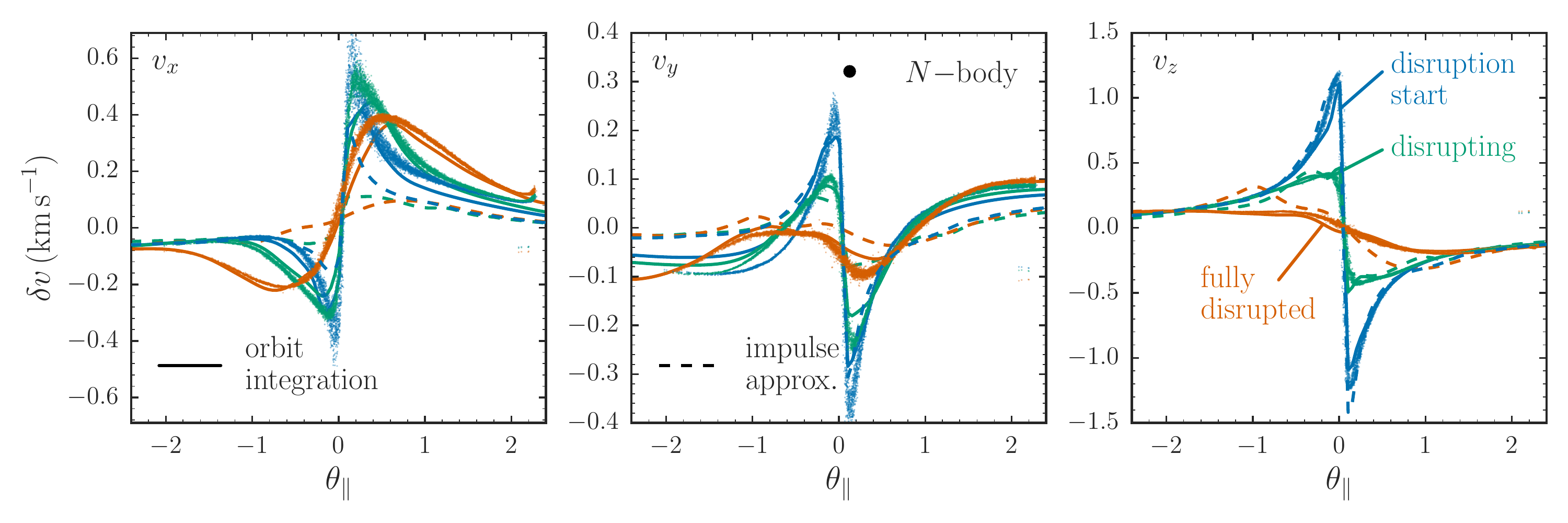}
\caption{Velocity kicks computed from the three $N$-body simulations
  shown in \figurename~\ref{fig:nbody}---`disruption start' from the
  left panel, `disrupting' from the middle panel, and `fully
  disrupted' from the right panel---represented as dots compared to
  two approximations. The kicks are shown as a function of the
  parallel angle coordinate $\theta_\parallel$ along the stream with
  respect to the impact point in action-angle coordinates
  \citep{Bovy14a}; the range in $\theta_\parallel$ shown spans almost
  the entire trailing arm of the stellar stream. The breakdown of the
  impulse approximation for the diffuse stream and for all streams in
  $v_x$ demonstrates that the full orbital path of the stream is
  responsible for the observed kicks.\label{fig:kicks}}
\end{figure*}

I compute the velocity kicks in the $N$-body simulations by
backwards-orbit-integration using \texttt{galpy} \citep{BovyGalpy} of
the stellar-stream particles after the interaction in the host
potential and comparing the velocity with that of the particles in a
simulation of the stellar stream with the same initial conditions, but
without the dark-mater subhalo. These velocity kicks are displayed in
\figurename~\ref{fig:kicks} as a function of angle along the
stream. Compared to the kicks from an interaction with a surviving
subhalo on the same orbit as the dark-matter stream, which peak at
$\approx(0.4,0.4,1.8)\kms$ in $(v_x,v_y,v_z)$ and are approximately
zero by $|\theta_\parallel| = 1$, it is clear that the kicks are
smaller and act over a more extended part of the stream, in agreement
with the considerations based on the impulse approximation
above. Interestingly, the kicks in $v_x$ are larger than that for the
surviving-subhalo interaction, with a similar amplitude for streams of
different lengths.

To understand the dynamics in the $N$-body simulation further, I
estimate the amount of stream mass passing through the impact point as
a function of time, by analyzing the dark-matter stream in
action-angle coordinates (cf. Ref. \citep{Bovy14a}). I then compute
the kicks using the impulse approximation above, accounting for the
movement of the stellar stream during the interaction. The motions of
the particles in the most diffuse stream are consistent with being
test particles in the host potential, but for the dark-matter streams
that are still in the process of tidal disruption I add a small
contribution from a single Plummer sphere to represent the remnant
subhalo. The resulting kicks are displayed as dashed lines in
\figurename~\ref{fig:kicks}. While the impulse approximation works
well for $v_y$ and $v_z$ for the shorter two streams, it fails for the
most diffuse stream and for all streams for $v_x$. The solid lines
show kicks computed by representing each dark-matter stream with a
random subsample of 300 particles, modeled as Plummer spheres with
$r_s = 10\pc$ and computing the kicks from each of these 300
interactions independently using orbit integration in the host+Plummer
potential. For the two shortest streams I again add small
contributions from a subhalo remnant. It is clear that this
approximation to the kicks matches the full $N$-body kicks in all
dimensions well, even at large offsets from the impact point. This
demonstrates that the impulse approximation breaks down because the
orbital motion of the dark-matter stream is important, rather than due
to the non-linear contributions from different parts of the stream.

Dark-matter subhalos are more realistically represented as NFW spheres
\citep{Navarro97a} rather than Plummer spheres. To determine whether
the effects discussed above are different for NFW halos, I have
repeated the simulations above, but modeling the dark-matter halos as
NFW halos with $M = 10^8\msun$, $r_s = 900\pc$, and a tidal truncation
radius of 2 kpc (chosen to be similar to subhalos in the Via Lactea-2
simulation \citep{Diemand08a} in the mass and radial range considered
here). The particle data for this NFW halo is sampled using the method
of Ref. \citep{McMillan07a}. These NFW DM halos disrupt and form tidal
tails of almost the same length and width as those in the Plummer
simulation above, and the effect on the GC stream is qualitatively the
same.

\emph{Discussion}---Stellar streams within tens of kpc from the
Galactic center typically encounter a few subhalos with masses of
$10^8$ to $10^9\msun$ \citep{Yoon11a}. Many of these may be in the
process of tidal disruption and give rise to velocity kicks along the
stellar streams similar to those in \figurename s~\ref{fig:straight}
and \ref{fig:kicks}. These kicks affect a larger part of the stream
and are slightly lower in amplitude. In standard analyses of the
impact of subhalos on stellar streams \citep{Erkal15b,Sanders15a},
both of these effects will lead to inferred $(M,r_s)$ with
anomalously-low concentrations compared to the cold-dark-matter
prediction. This will be the telltale sign that the stellar stream has
been hit by a dark-matter stream rather than a surviving subhalo. From
the $N$-body simulations above, diffuse streams can give substantial
kicks for at least $\approx0.5\Gyr$, so the probability of catching a
dark-matter halo in the act of disrupting is high.

Analyses of the kinematics of stellar streams (cf. \citep{Erkal15b})
can therefore determine the prevalence of dark-matter streams in the
Milky Way halo. Many additional stellar streams within tens of kpc are
expected to be found soon using data from \emph{Gaia}
\citep{Perryman01a} and we will therefore soon have plenty of
potential targets for a dark-matter-stream search. Such a measurement
would have profound implications for dark-matter direct-detection
experiments \citep{Kuhlen10a} and would provide an important
constraint on the formation of halos in the hierarchical cosmological
framework.

The $N$-body simulations above demonstrate that stellar streams are
uniquely sensitive to the full orbital path of dark-matter
streams. This is unlike the case of subhalo--stream interactions,
which are typically well modeled using the impulse approximation. In
this approximation, the velocity kicks remain the same when the mass
of the perturber and the relative fly-by velocity are changed by the
same factor \citep{Erkal15b}. Computing the kicks for a simulation
like the ``disrupting'' case in \figurename~\ref{fig:kicks}, but with
the mass and relative fly-by velocity scaled down by half, I find
velocity kicks that are different in amplitude and width by $50\,\%$,
while these can be measured to $\approx10\,\%$ from \emph{Gaia} and
\emph{LSST} data (cf.  Ref. \citep{Erkal15b}), and the full orbit of
the dark-matter stream could thus be precisely constrained. If a
likely dark-matter stream from a recently disrupted subhalo is
detected as discussed above, detailed observations of the kinematics
along the stellar stream may therefore reveal the full orbital path
and present position of a dark subhalo. Such an object would be a
tantalizing target for indirect dark-matter detection experiments.

\begin{acknowledgments}
 It is my pleasure to thank Neal Dalal for conversations that inspired
 this work and the anonymous referees for helpful comments that
 improved this paper. This research received financial support from
 the Natural Sciences and Engineering Research Council of Canada.
\end{acknowledgments}

\bibliography{ms.bib}

\end{document}